\documentclass{iau}
\usepackage{graphicx}

\title[DIBs in (proto-) fullerene-rich environments] 
{Diffuse interstellar bands in (proto-) fullerene-rich environments}

\author[D. A. Garc\'{\i}a-Hern\'andez]   
{D. A. Garc\'{\i}a-Hern\'andez$^{1,2}$
\thanks{D.A.G.H. acknowledges support provided by the Spanish
Ministry of Economy and Competitiveness under grant AYA$-$2011$-$29060.}
}

\affiliation{$^1$Instituto de Astrof\'{\i}sica de Canarias, \\ C/ Via L\'actea s/n,
E$-$38200 La Laguna, Tenerife, Spain \\ email: {\tt agarcia@iac.es} \\[\affilskip]
$^2$Departamento de Astrof\'{\i}sica, Universidad de La Laguna (ULL), \\
E$-$38206 La Laguna, Tenerife, Spain
}

\pubyear{2013}
\volume{297}  
\pagerange{119--126}
\jname{The Diffuse Insterstellar Bands}
\editors{Jan Cami \& Nick Cox, eds.}
\begin{document}

\maketitle

\begin{abstract}
The recent infrared detection of fullerenes (C$_{60}$ and C$_{70}$) in
Planetary Nebulae (PNe) and R Coronae Borealis (RCB) stars offers a beautiful
opportunity for studying the diffuse interstellar bands (DIBs) in sources where
fullerenes are abundant. Here we present for the first time a detailed
inspection of the optical spectra of the hot RCB star DY Cen and two fullerene
PNe (Tc 1 and M 1-20), which permits us to directly explore the fullerenes -
DIB connection. The DIB spectrum of DY Cen (Garc\'{\i}a-Hern\'andez et al.
2012a) is remarkably different from that in fullerene PNe
(Garc\'{\i}a-Hern\'andez \& D\'{\i}az-Luis 2013). In particular, Tc 1 displays
unusually strong 4428 \AA~ and 6309 \AA~DIBs, which are normal (or not seen) in
DY Cen. On the other hand, DY Cen displays an unusually strong 6284 \AA~DIB that
is found to be normal in fullerene PNe. We also report the detection of new
broad and unidentified features centered at 4000 \AA~and 6525 \AA~in DY Cen and
Tc 1, respectively. We suggest that the new 4000 \AA~band seen in DY Cen may be
related to the circumstellar proto-fullerenes seen at infrared wavelengths
(Garc\'{\i}a-Hern\'andez et al. 2012a). However, the intense 4428 \AA~DIB
(probably also the 6309 \AA~DIB and the new 6525 \AA~band) may be related to the
presence of larger fullerenes (e.g., C$_{80}$, C$_{240}$, C$_{320}$, and
C$_{540}$) and buckyonions (multishell fullerenes such as C$_{60}$@C$_{240}$ and
C$_{60}$@C$_{240}$@C$_{540}$) in the circumstellar envelope of Tc 1
(Garc\'{\i}a-Hern\'andez \& D\'{\i}az-Luis 2013).
\end{abstract}

\firstsection 
\section{Fullerenes in RCB stars and PNe}

Fullerenes (e.g., C$_{60}$, C$_{70}$) are very stable molecules that are very
important for interstellar/circumstellar chemistry and that may explain many
astrophysical phenomena such as the mysterious diffuse interstellar bands (DIBs)
and the intense UV bump at 2170 \AA~(e.g., Iglesias-Groth 2007; Cataldo \&
Iglesias-Groth 2009). Fullerenes were discovered in the laboratory (Kroto et al.
1985) and they have been found on Earth and on meteorites. Indeed, the 9577 and
9632 \AA\ DIBs observed in some reddened stars lie near two electronic
transitions of C$_{60}$$^{+}$ (Foing \& Ehrenfreund 1994; see also Bern\'e et
al. 2013 for a recent IR detection of the C$_{60}$ cation). At laboratory,
fullerenes are efficiently produced under H-poor conditions (e.g., Kroto et al.
1985) and the so-called R Coronae Borealis (RCB) stars (extremely H-deficient
stars) were thus expected to efficiently produce fullerenes (Goeres \& Sedlmayr
1992). In 2010, Garc\'{\i}a-Hern\'andez, Rao \& Lambert looked for C$_{60}$ in a
complete sample of about 30 RCBs by using the Spitzer Space Telescope. We got
the unexpected result that C$_{60}$-like mid-IR features were only detected in
those RCBs with some H (Garc\'{\i}a-Hern\'andez et al. 2011a). In particular,
C$_{60}$-like emission features were detected in the least H-deficient RCBs DY
Cen and V854 Cen, which also show strong polycyclic aromatic hydrocarbon (PAH)
features. Because of the unexpected result in RCBs, then we looked for
fullerenes in $\sim$240 Planetary Nebulae (PNe) by using data from our own
Spitzer projects; five clear fullerene detections were found. Meanwhile, Cami et
al. (2010) reported the extraordinary discovery of the first IR detection of
C$_{60}$ and C$_{70}$ fullerenes in the young PN Tc 1. At the same time,
Garc\'{\i}a-Hern\'andez  et al. (2010) confirmed the surprising results obtained
in RCBs, showing that all PNe with fullerenes (including Tc 1) are low-mass
C-rich PNe with normal H abundances. Contrary to the RCB stars, the Spitzer PNe
spectra display C$_{60}$-like features in conjunction with very weak PAHs. This
challenged our understanding of the fullerenes formation in space, showing that,
contrary to general expectation, fullerenes are efficiently formed in H-rich
circumstellar environments only. Furthermore, the detection of fullerenes in RCB
stars and PNe suggests that large fullerenes may be formed as decomposition
products of hydrogenated amorphous carbon (HAC) dust (Garc\'{\i}a-Hern\'andez et
al. 2010, 2011a,b, 2012b; Bernard-Salas et al. 2012; Micelotta et al. 2012).

\begin{figure}[b]
\begin{center}
 \includegraphics[angle=0,scale=.3]{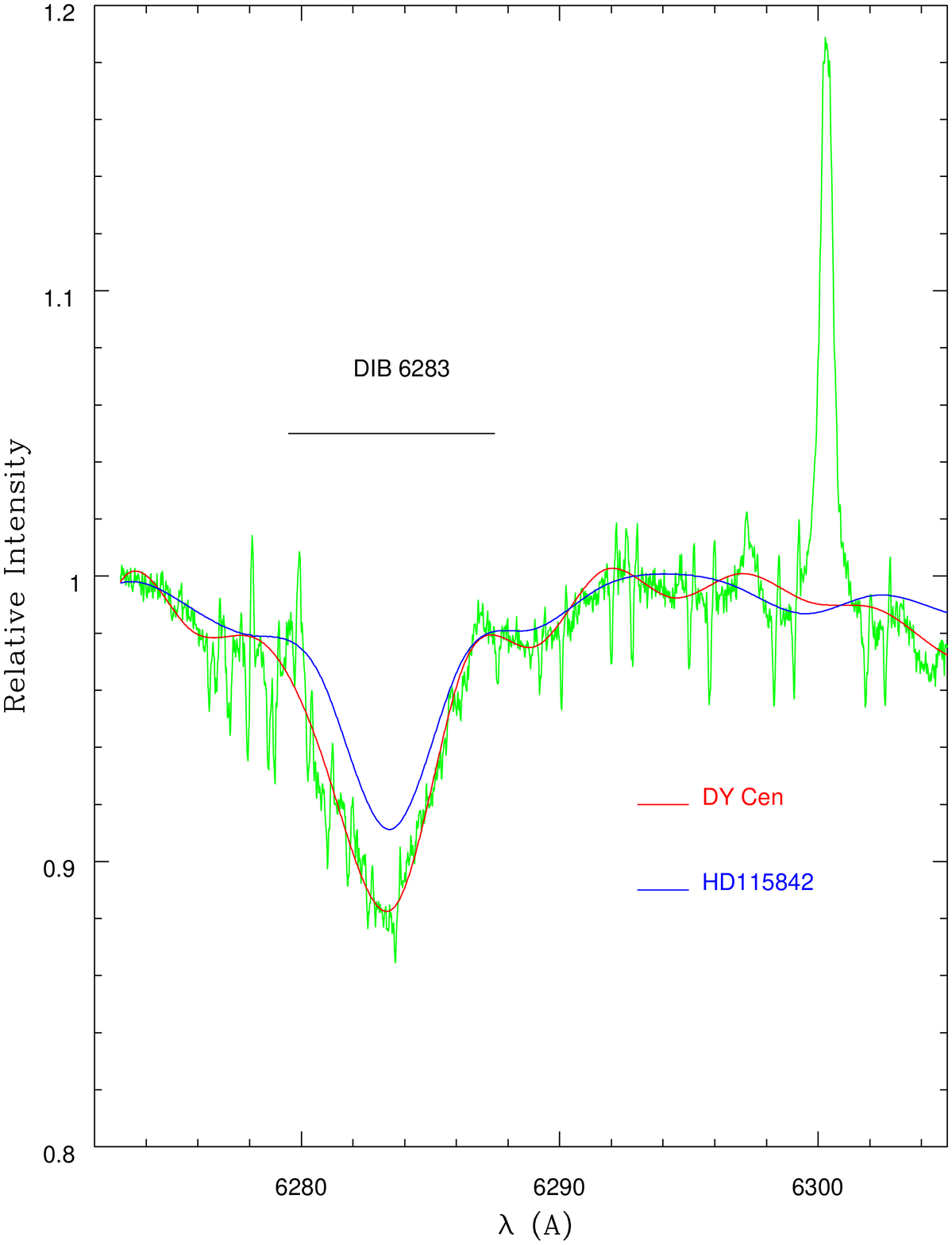}%
  \includegraphics[angle=0,scale=.3]{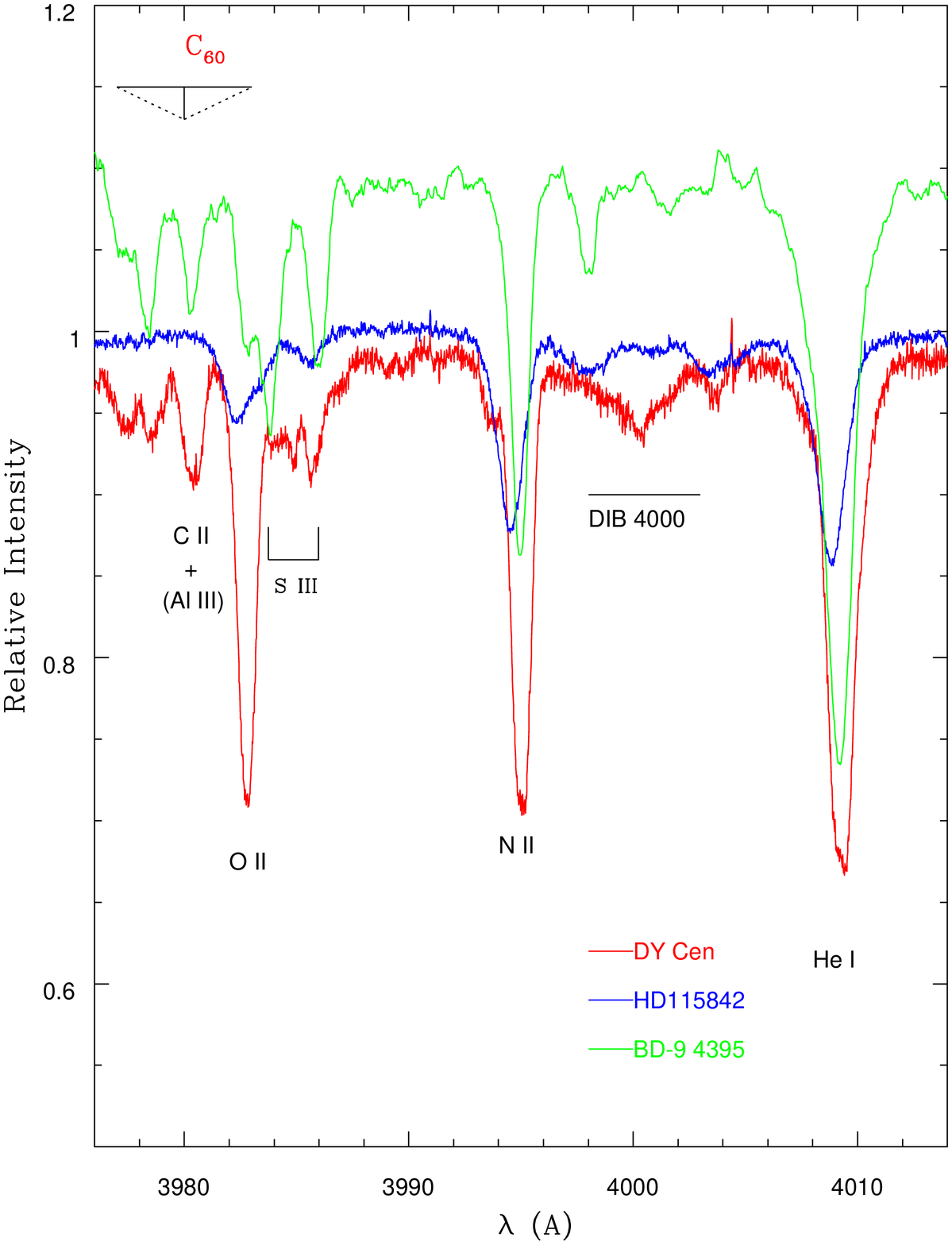}
 \caption{Left panel: Spectral region around the 6284\AA~DIB in DY Cen (in red)
and HD 115842 (in blue). The telluric line corrected spectrum of both stars are
shown by smooth lines (red and blue). The non telluric corrected spectrum in DY
Cen is also shown (in green). Right panel: The spectra of DY Cen (in red) and HD
115842 (in blue) around and 4000 \AA~where the eHe star BD $-9^\circ$ 4395 is
also displayed (in green). The expected position and FWHM of the C$_{60}$
feature at 3980 \AA~are marked on top of the spectra; there is no evidence
(additional absorption) in DY Cen for the presence of this neutral C$_{60}$
feature. However, there is an additional absorption band at 4000 \AA\ in DY Cen,
which is not present in HD 115842 neither in BD $-9^\circ$ 4395.}
   \label{fig1}
\end{center}
\end{figure}

\begin{figure}[b]
\begin{center}
 \includegraphics[angle=0,scale=.3]{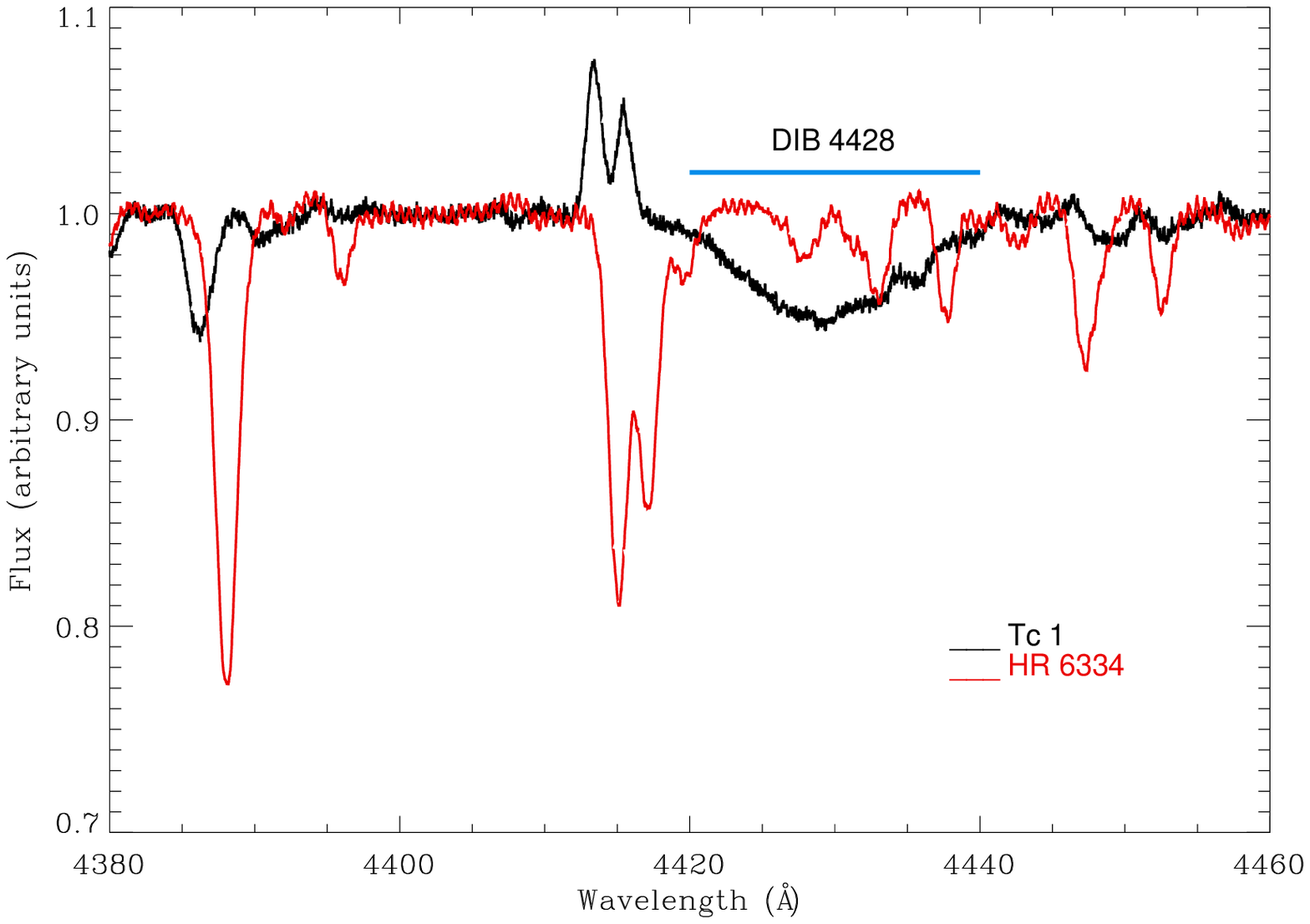}%
  \includegraphics[angle=0,scale=.3]{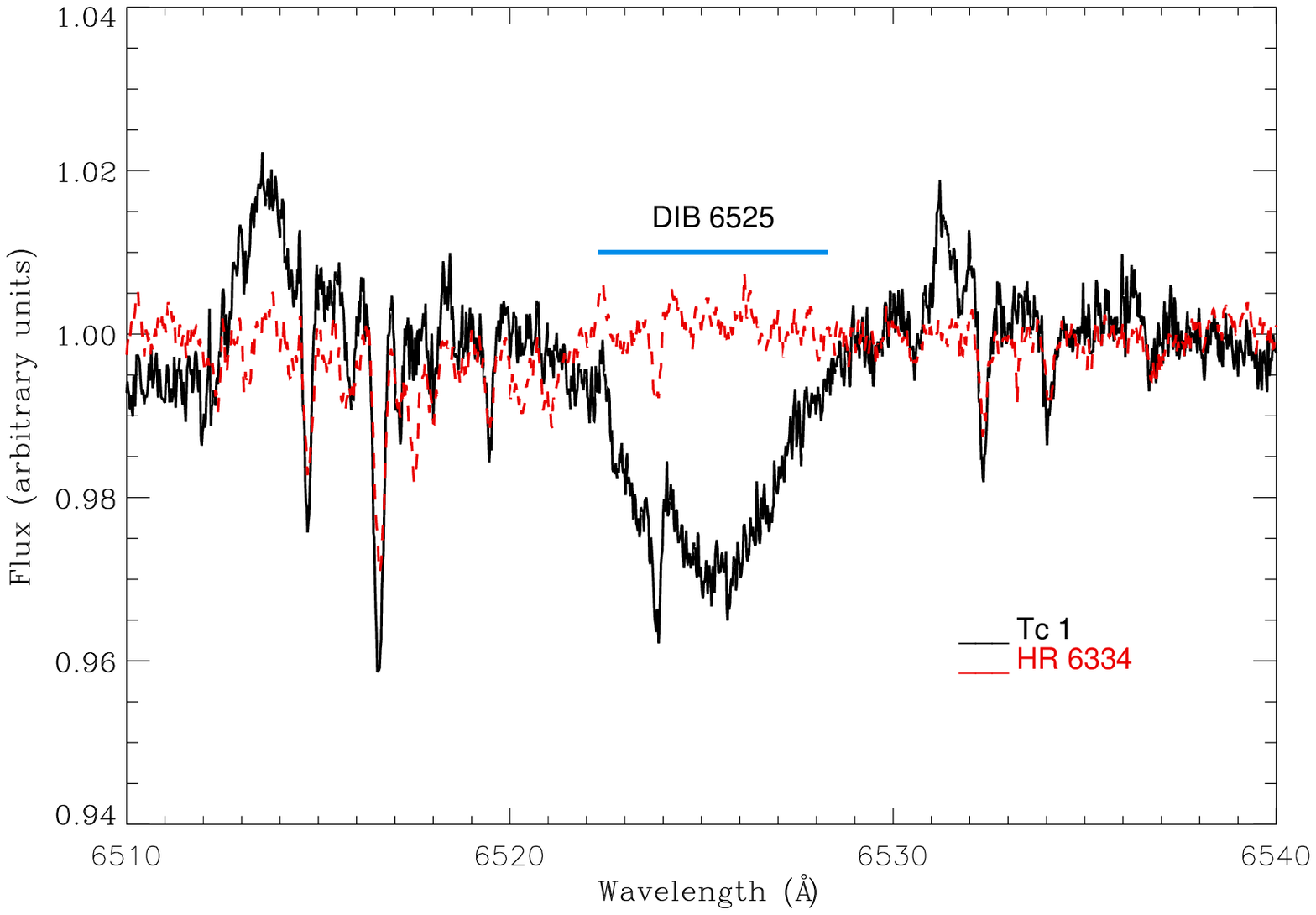}
 \caption{Left panel: Spectra of Tc 1 (in black) and HR 6334 (in red) around
the  4428 \AA\ DIB. This DIB is found to be unusually strong in Tc 1 while HR
6334 - with a higher E(B-V) of 0.42 - does not show evidence of its presence.
Right panel: Spectral region around the new broad unidentified band at 6525 \AA\
in Tc 1 (in black) and HR 6334 (in red).}
   \label{fig2}
\end{center}
\end{figure}

Thus, fullerenes and related large C-based molecules (e.g., other fullerenes as
stable exohedral or endohedral metallo-complexes) might be ubiquitous in the
interstellar medium and continue to be serious candidates for the DIB carriers.
A detailed analysis of the DIBs towards fullerene-containing - accompanied or
not by PAH molecules - space environments is very useful to learn about the
nature of the DIB carriers. In this context, the recent detections of large
fullerene-like molecules in RCBs and PNe offer a beautiful opportunity for
studying the DIB spectrum of sources where fullerenes and fullerene-related
molecules are abundant. Here we present for the first time a detailed inspection
of the optical spectra of the hot RCB star DY Cen and two fullerene PNe (Tc 1
and M 1-20), which permits us to directly explore the fullerenes - DIB
connection.  

\section{Our DIB's survey}

We acquired high-resolution (R$\geq$15,000) and high signal-to-noise
(S/N$\geq$200) VLT-UVES optical ($\sim$3300-9450 \AA) spectra of the RCB star DY
Cen and PN Tc 1. The PN M 1-20 was also observed although at lower S/N. Nearby
B-type comparison stars were observed on the same dates as the corresponding
science objects using the same VLT-UVES set-up. The observational details are
not repeated here and we refer the reader to Garc\'{\i}a-Hern\'andez et al.
(2012a) and Garc\'{\i}a-Hern\'andez \& D\'{\i}az-Luis (2013) for the
observations of DY Cen and Tc 1 (and M 1-20), respectively. The Spitzer IR
spectrum of DY Cen is dominated by PAH-like features with weaker C$_{60}$-like
features (Garc\'{\i}a-Hern\'andez et al. 2011a) while Tc 1 displays a
C$_{60}$-dominated IR spectrum with very weak (or absent) PAH bands (Cami et al.
2010; Garc\'{\i}a-Hern\'andez et al. 2010). The goal of our optical observations
is to study the characteristics of DIBs in fullerene-containing environments as
well as to shed some light about the possible fullerenes - DIB connection. 

We find that the classical and well-studied DIBs (e.g., those at 4428, 5780,
5797, 5850, 6196, 6379, and 6614 \AA) towards DY Cen are normal for its
reddening. The only exception is the DIB at 6284 \AA\ (possibly also the
7223\AA\ DIB) (see Garc\'{\i}a-Hern\'andez et al. 2012a for more details).
Figure 1 (left panel) shows the region of 6284 \AA\ for DY Cen and the nearby
comparison star HD 115824. It is clear that this DIB towards DY Cen is stronger
than towards HD 115842, suggesting that the carrier of the 6284\AA\ DIB (along
with 7223 \AA) is different from the rest of the classical DIBs. Also
interesting is that we detect in DY Cen a broad (FWHM$\sim$2 \AA) unidentified
feature centered at $\sim$4000 \AA, which is seen in DY Cen only (Fig. 1; right
panel). Note that no DIBs are known at this wavelength (see e.g., Hobbs et al.
2008) and no molecule is known to exhibit a strong electronic transition at
$\sim$4000 \AA. In addition, Garc\'{\i}a-Hern\'andez et al. (2012a) have
reported the non-detection of the strongest C$_{60}$ electronic transitions
(e.g., those at $\sim$3760, 3980, and 4024 \AA) in DY Cen (see Fig. 1, right
panel); C$_{60}$ IR column density estimates are 1000 times higher than the
optical detection limits.

Similarly to DY Cen, we find the strongest DIBs (e.g., those at 5780, 5797,
5850, 6196, 6270, 6284, 6380, and 6614 \AA) most commonly found in the ISM to be
normal in Tc 1 and M 1-20 (see Garc\'{\i}a-Hern\'andez \& D\'{\i}az-Luis 2013
for more details). This may suggest that the carriers of the latter well-studied
DIBs are not particularly overabundant in fullerene PNe. Surprisingly, the
well-studied DIB at 4428 \AA~as well as the weaker 6309\AA\ DIB (see e.g., Hobbs
et al. 2008) are found to be unusually strong towards Tc 1; the 4428\AA~DIB is
unusually strong in M 1-20 too. Figure 2 (left panel) compares the 4428 \AA~DIB
in Tc 1 with that in the nearby comparison star HR 6334. Adopting a Lorentzian
profile for the 4428 \AA\ DIB, we obtain an EQW of 860 m\AA, which is at least a
factor of two greater than expected from the reddening in Tc 1. An unidentified
broad (FWHM $\sim$5 \AA) feature at 6525 \AA\ is also detected in Tc
1\footnote{A few other unidentified  bands and/or unusually strong DIBs seem to
be present in Tc 1 (see Manchado et al. these proceedings).}. Figure 2 (right
panel) displays the unidentified 6525\AA\ band in the Tc 1 spectrum in
comparison with the star HR 6334. Again, the most intense optical bands of 
neutral C$_{60}$ are lacking in the Tc 1 spectrum; although this could be
explained by the low C$_{60}$ column density estimated from the C$_{60}$ IR
features if the neutral C$_{60}$ emission peaks far away from the central star
(Garc\'{\i}a-Hern\'andez \& D\'{\i}az-Luis 2013). Finally, the DY Cen's
unidentified  4000 \AA~band is not seen in Tc 1.

\section{Diffuse interstellar bands in (proto-) fullerene-rich environments}

Table 1 compares the DIBs seen in the RCB star DY Cen with those in the
fullerene PN Tc 1. This table naturally prompts the question of why DIBs are so
different in the fullerene-containing environments around RCB stars and PNe? 

Based on their laboratory spectroscopy of HAC nanoparticles, Duley \& Hu (2012)
propose that the C$_{60}$ IR features seen in sources with PAH-like dominated IR
spectra such as DY Cen are attributable to proto-fullerenes or fullerene
precursors rather than to C$_{60}$. Our non-detection of neutral C$_{60}$ in the
DY Cen optical spectrum may support the Duley \& Hu (2012) HAC laboratory
results. Thus, the new 4000\AA~band seen in DY Cen may be related to the
circumstellar proto-fullerenes; perhaps an organic compound containing
pentagonal rings. These pentagonal carbon rings are usually present in HAC
nanoparticles and nanotubes, suggesting that they may be intimately related with
the formation process of fullerenes. In addition, the 4428 \AA\ DIB has been
linked to fullerenes bigger than C$_{60}$ and/or buckyonions such as
C$_{60}$@C$_{240}$ and C$_{60}$@C$_{240}$@C$_{540}$ (Iglesias-Groth 2007). Our
findings in DY Cen would be consistent with fullerenes and fullerenes-containing
molecules not being especially overabundant towards DY Cen. 

On the other hand, the non detection of neutral C$_{60}$ in the Tc 1 optical
spectrum is intriguing. Tc 1 - with no or very weak PAHs - is expected to be
rich in C$_{60}$  (Duley \& Hu 2012). An exotic explanation may be that larger
fullerenes or more complex fullerene-based molecules are present. This is
suggested by the unusually strong 4428\AA~DIB. Photo-absorption theoretical
models of large fullerenes (C$_{80}$,  C$_{240}$, C$_{540}$) and buckyonions
(C$_{60}$@C$_{240}$, C$_{60}$@C$_{240}$@C$_{540}$) (Iglesias-Groth 2007) display
strong transitions around 4428 \AA. In this framework, the 4428\AA~DIB may be
explained by the transitions (superposition) of fullerenes bigger than C$_{60}$
and multishell fullerenes (buckyonions) (Iglesias-Groth 2007). Recent studies of
the 4428\AA~DIB also suggest the carrier to be a resistant,  large and compact
neutral molecule (van Loon et al. 2013). 

\begin{table}
  \begin{center}
  \caption{Overview of DIBs in fullerene-containing RCBs and PNe.}
  \label{tab1}
 {\scriptsize
  \begin{tabular}{|l|c|c|}\hline 
{\bf DIB} & {\bf RCBs} & {\bf PNe}  \\ 
   &  (DY Cen) & (Tc 1) \\ \hline
4000 \AA\   &  yes & no \\ 
4428 \AA\   &  normal & strong \\ 
6284 \AA\   & strong &  normal \\
6309 \AA\   &  no  & yes \\ 
6525 \AA\   &  no  & yes \\ 
\hline
  \end{tabular}
  }
 \end{center}
\end{table}

\section{Concluding remarks}

The IR detection of C$_{60}$ in RCB stars and PNe offers the opportunity of
studying DIBs in environments where fullerenes are abundant. We have shown here
that DIBs in RCB stars and PNe are remarkably different. The new 4000 \AA~band
detected in the RCB star DY Cen is suggested to be related with proto-fullerenes
or fullerene-precursors (Garc\'{\i}a-Hern\'andez et al. 2012a). However, the
unusually strong 4428\AA~DIB (probably also the weaker 6309\AA~DIB and the
unidentified 6525\AA~band) in PNe is suggested to be related with the presence
of large fullerenes and buckyonions (Garc\'{\i}a-Hern\'andez \& D\'{\i}az-Luis
2013) as previously pointed out by Iglesias-Groth (2007) from theoretical
considerations. 

At present, the HAC's photochemical processing seems to be the most promising
C$_{60}$ formation route; at least in the complex circumstellar envelopes of RCB
stars and PNe (see e.g., Garc\'{\i}a-Hern\'andez et al. 2012b; and references
therein) but see also Bern\'e \& Tielens (2012). Larger fullerenes and
fullerene-based molecules may form from pre-existing C$_{60}$ molecules (e.g.,
Dunk et al. 2012), opening the possibility of forming a rich family of
fullerene-related molecules such as buckyonions, metallofullerenes, and
fullerene adducts. These complex fullerene-based molecules may emit through the
same IR vibrational modes (e.g., as isolated C$_{60}$ and C$_{70}$), being
indistinguishable from C$_{60}$ (and C$_{70}$) on the basis of IR spectra alone.
In particular, fullerenes and PAHs may be mixed in the circumstellar envelopes
of fullerene PNe (e.g., M 1-20) and fullerene/PAH adducts may form via
Dies-Alder cycloaddition reactions (Garc\'{\i}a-Hern\'andez et al. 2013).
Indeed, very recent laboratory work demonstrates that fullerene/PAH adducts -
such as C$_{60}$/anthracene Diels-Alder adducts - display mid-IR features
strikingly coincident with those from C$_{60}$ and C$_{70}$
(Garc\'{\i}a-Hern\'andez et al. 2013).

\end{document}